\documentclass[%
 reprint,
 aps,
 pre,
 superscriptaddress,
]{revtex4-2}

\usepackage{graphicx}% Include figure files
\usepackage{dcolumn}% Align table columns on decimal point
\usepackage{bm}% bold math
\usepackage[hypertexnames=false]{hyperref}% add hypertext capabilities
\usepackage{amsfonts}       % blackboard math symbols
\usepackage{nicefrac}       % compact symbols for 1/2, etc.
\usepackage{microtype}      % microtypography
\usepackage{amsmath}       % blackboard math symbols
\usepackage{amssymb}       
\usepackage{amsthm}
\usepackage{mathtools}
\usepackage{blindtext}
\usepackage{centernot}
\usepackage{algorithm}
\usepackage{algorithmic}
\usepackage{subfigure}
\usepackage{svg}
\usepackage{bbm}
\usepackage{tikz}
\usepackage{tikz-cd}
\usetikzlibrary{positioning}
\usetikzlibrary{calc}

\DeclareFontFamily{U}{mathx}{\hyphenchar\font45}
\DeclareFontShape{U}{mathx}{m}{n}{
      <5> <6> <7> <8> <9> <10>
      <10.95> <12> <14.4> <17.28> <20.74> <24.88>
      mathx10
      }{}
\DeclareSymbolFont{mathx}{U}{mathx}{m}{n}
\DeclareFontSubstitution{U}{mathx}{m}{n}
\DeclareMathAccent{\widecheck}{0}{mathx}{"71}

\newcommand{\graphsum}[2]{\psi\left(#1, #2\right)}

\graphicspath{{fig/}}

\begin{document}

\preprint{APS/123-QED}

\title{Motif-based mean-field approximation of interacting particles on clustered networks}

\author{Kai~Cui\hspace{0.5mm}\href{https://orcid.org/0000-0002-2605-0386}{\includegraphics[width=3mm]{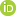}}}
 \email{kai.cui@bcs.tu-darmstadt.de}
\affiliation{%
  Technische Universität Darmstadt, Germany
}%

% \collaboration{CLEO Collaboration}%\noaffiliation

\author{Wasiur~R.~KhudaBukhsh\hspace{0.5mm}\href{https://orcid.org/0000-0003-1803-0470}{\includegraphics[width=3mm]{ORCID-iD_icon-16x16}}}
 \email{wasiur.khudabukhsh@nottingham.ac.uk}
%  \altaffiliation[Also at]{School of Mathematical Sciences, University of Nottingham}%Lines break automatically or can be forced with \\
%  \homepage{http://www.Second.institution.edu/~Charlie.Author}
\affiliation{
 University of Nottingham, United Kingdom
}%

\author{Heinz~Koeppl\hspace{0.5mm}\href{https://orcid.org/0000-0002-8305-9379}{\includegraphics[width=3mm]{ORCID-iD_icon-16x16}}}%
 \email{heinz.koeppl@bcs.tu-darmstadt.de}
\affiliation{%
  Technische Universität Darmstadt, Germany
}%

\date{\today}

\allowdisplaybreaks

\begin{abstract}
    Interacting particles on graphs are routinely used to study magnetic behaviour in physics, disease spread in epidemiology, and opinion dynamics in social sciences. The literature on mean-field approximations of such systems for large graphs typically remains limited to specific dynamics, or assumes cluster-free graphs for which standard approximations based on degrees and pairs are often reasonably accurate. Here, we propose a motif-based mean-field approximation that considers higher-order subgraph structures in large clustered graphs. Numerically, our equations agree with stochastic simulations where existing methods fail. 
\end{abstract}

\maketitle

%\tableofcontents

% \paragraph{Introduction}
With applications in as disparate branches of science as statistical physics \citep{glauber1963time}, epidemiology \citep{kiss2017mathematics,Sahneh2013Generalized,Pastor-Satorras2001Scalefree,Pastor-Satorras2015Complex, PhysRevX.4.041005, PhysRevResearch.2.033005, PhysRevLett.126.118301, PhysRevLett.117.028302}, chemistry and systems biology \citep{Anderson2020Prevalence,Anderson2021Deficiency}, social science \citep{Kashin2021Opinion,Juul2019Hipsters}, and computer science \citep{bruneo2012markovian, gribaudo2008analysis, VanMieghem2009Virus}, interacting particles on complex networks constitute an important class of models in the mathematician's and physicist's toolkit \citep{albert2002statistical,barrat2008dynamical,durrett_2006_RGD,hofstad2017_RGCN}. They describe systems where individual entities (particles), endowed with local states, interact with a subset of other entities (neighbors) and transition from one state to another as time evolves. For instance, in epidemiology the local state space consists of immunological statuses, such as susceptible, infected, removed etc. Who interacts with whom defines a graph with the particles as the vertices.

The time evolution of the ensemble of particle states is often described by a continuous-time Markov jump process, for which discrete-time analysis can be insufficient \citep{PhysRevE.94.052125}. As the number of particles increases, the exponentially growing combinatorial state space renders exact stochastic analysis prohibitive. To this end, the standard mean-field theoretic approach has been to describe the non-equilibrium dynamics of interacting particle systems via Ordinary Differential Equations (ODEs) for the proportions of particles in each state. Together with control \citep{lasry2007mean}, learning-based methods \citep{guo2019learning, cui2021approximately} and graph limit theory \citep{Lovasz2012LargeNetworks, caines2019graphon, bayraktar2020graphon, cui2021learning}, mean-field models can enable analysis of otherwise intractable settings \citep{laguzet2015individual, djehiche2017mean, perrin2021mean}. More advanced mean-field approximations, such as heterogeneous mean-fields \citep{sood2005voter}, pair approximations \citep{doi:https://doi.org/10.1002/9781444311501.ch4, vazquez2008analytical, pugliese2009heterogeneous, PhysRevResearch.1.033024, Pellis2015Pairwise} or approximate master equations (AME) \citep{gleeson2011high, gleeson2013binary, doi:10.1137/16M1109345, lindquist2011effective} and extensions thereof \citep{peralta2018system, PhysRevResearch.2.043370, PhysRevLett.119.208301, PhysRevLett.116.258301}, acknowledge the heterogeneity of the particles' behaviors due to the graph structure and incorporate vertices' degrees and edge counts (pairs). Though they provide reasonable accuracy for a number of applications, they are generally not asymptotically exact in that they do not agree with the Functional Law of Large Numbers (FLLN) limits of the corresponding stochastic processes, agreeing only in certain special cases \citep{Jacobsen2018LargeGraph,KhudaBukhsh2022FCLT}. Even for calculations of critical parameter values, standard mean-field approximations are often inaccurate \citep{Chatterjee2009Contact}. Nevertheless, their simplicity and intuitiveness have commonly justified mean-field approaches despite their inexactness.

In this paper, we propose a simple and elegant derivation of a general motif-based mean-field approximation for interacting particles on bounded-degree graphs to address two crucial shortcomings of the state-of-the-art: (i) The implicit assumption of cluster-free graphs \citep{PhysRevE.85.026106}. In practice, graphs encountered are far from cluster-free and exhibit complex structures \citep{bruneo2012markovian, battiston2020networks, PhysRevLett.127.158301, PhysRevLett.126.098301} (e.g., neural and transportation networks \citep{doi:10.1126/science.aad9029}), which greatly affect e.g. cascades in correlated networks \citep{PhysRevE.77.046117}. Here, we go beyond correlation coefficients and account for arbitrary subgraph structures called motifs \citep{Schwartze2021Motif} beyond standard degree and edge-based calculations. (ii) The restriction to special cases (e.g. SIR epidemics, \citep{ritchie2016beyond}) or dynamics driven by simple neighborhood counts. For instance, infection rates are often assumed to depend only on the number of infected neighbors, while in practice shared connections among neighbors and the shape of the induced neighborhood subgraph are too important to neglect (e.g., simplicial dynamics \citep{bruneo2012markovian, PhysRevE.103.032301}). Though there exist a multitude of works on the analysis of clustered graphs \cite{PhysRevE.85.026106, PhysRevE.54.2351, vlasov2017hub, PhysRevLett.94.018106, PhysRevE.68.046126}, to the best of our knowledge, we provide the first general approximation that takes into account both of these aspects into a single coherent mean-field framework. We now introduce the mathematical model before explaining how our approximation addresses the above two issues.

\paragraph{Model}

A convenient way of generating random graphs is via the Configuration Model (CM) \citep{hofstad2017_RGCN,newman2018networks}, which allows specifying either a degree sequence or probability law from which the degrees are sampled. Each vertex is assigned as many half-edges as its degree. We may need to add or drop a parity edge if the degree sequence is not graphical, but its contribution is negligible in large graphs \citep[Section 7.6, pp. 239]{hofstad2017_RGCN}. The configuration model graph is then constructed by uniformly-at-random matching of all available half-edges. As $N$, the number of vertices, grows to infinity, the numbers of self-loops and multiple edges have independent Poisson limits whose means depend only on the first two moments of the degree distribution \citep[Theorem 3.1.2]{durrett_2006_RGD}. Therefore, their contributions to the limits of various counts scaled by $1/N$ (standard mean-field scaling) vanish in the limit.

\begin{figure}
    \center
    \includegraphics[width=\linewidth]{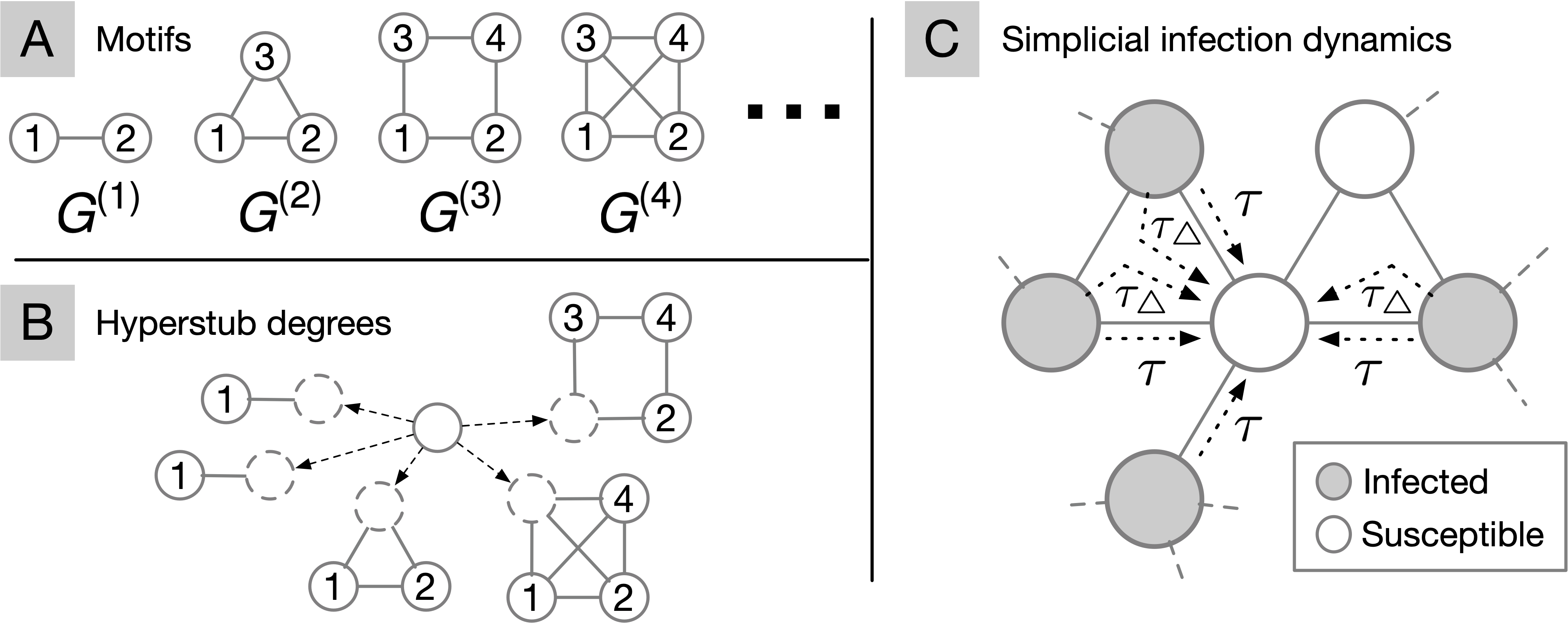}
    \caption{Schematic illustration of the model. \textbf{A:} A finite number of network motifs occurring in the network. \textbf{B:} An example ECM vertex with two motif participations as vertex $2$ in $G^{(1)}$ and one each as vertices $3, 1, 3$ in $G^{(2)}, G^{(3)}, G^{(4)}$ respectively, i.e. $d^1 \equiv (d^{1,1}, d^{1,2}) = (0, 2)$, $d^2 = (0,0,1)$, $d^3 = (1,0,0,0)$, $d^4 = (0,0,1,0)$. \textbf{C:} Simplicial SIS dynamics \citep{bruneo2012markovian} as an example of general neighborhood-dependent dynamics. Susceptible vertices are infected at rate $\tau$ by infected neighbors, and additionally at rate $\tau_\triangle$ for each shared neighbor.}
    \label{fig:explain1}
\end{figure}

To introduce higher-order structure, we adopt the Extended Configuration Model (ECM) \citep{PhysRevE.82.066118} -- also known as hyperstub configuration model \citep{ritchie2016beyond, ritchie2017generation}. Denoting vertices and edges of graphs $H$ by $V(H)$ and $E(H)$ respectively, and given $M$ graphical network motifs $G^{(1)}, \ldots, G^{(M)}$ with $N_1, \ldots, N_M$ vertices, we construct an ECM on $N$ vertices by specifying higher-order motif participation counts (hyperstub degrees) $(d_1, \ldots, d_N)$, where $d_v \equiv (d^1, \ldots, d^M) \in \mathcal D$, $d^i \equiv (d^{i,1}, \ldots, d^{i, N_i})$, and $d^{i,j} \in \mathbb N_0$ denotes the number of participations (hyperstubs) as the $j$-th vertex (role) in the motif $G^{(i)}$ (see Figure~\ref{fig:explain1}). As in the standard CM, hyperstubs are first generated for each node in accordance with a limiting hyperstub degree distribution $P(d)$. Subsequently, for each possible motif, we iteratively sample hyperstubs of each motif vertex role and add edges wherever the underlying motif has an edge, repeating until no hyperstubs are left. 

To describe the dynamics of the interacting system, we will consider time-indexed colored ECM graphs $\{G_t\}_{t \geq 0}$. Each vertex is endowed with a local (finite) state space $\mathcal{X}$. Denote the state of vertex $v$ at time $t$ -- interpreted as color -- as $G_t[v]$, and define the colored neighborhoods $\mathcal N_t^{(v)}$ as colored subgraphs of $G_t$ with fixed center vertex $v$, induced by the set of all vertices participating in motifs with $v$. Treated as a stochastic process,  $G_t$ is a Markov jump process with infinitesimal rates $\lambda^{x \to y}(\mathcal N_t^{(v)})$, depending on $v$ only via its colored neighborhood configuration, i.e. the rate
for vertex $v$ to jump from state $x$ to $y \; (\neq x)$ is given by
\begin{multline}
    \mathbb P(G_{t+h}[v] = y \mid G_{t}[v] = x, \mathcal N_t^{(v)}) \\
    = \lambda^{x \to y}(\mathcal N_t^{(v)}) h + o(h) \, .
\end{multline} 

Note that the rate functions $\lambda^{x \to y}$ depend on the entire subgraph and its coloring up to isomorphism (not only neighbor state counts), and therefore generalize those considered in standard mean-field approximations. To illustrate this, define the neighbor evaluation function
\begin{align}
    \psi(\mathcal N^{(v)}, f) \equiv \sum_{n \in V(\mathcal N^{(v)}) \colon (v, n) \in E(\mathcal N^{(v)})} f(\mathcal N^{(v)}[n])
\end{align} 
for any $f \colon \mathcal{X} \to \mathbb{R}$ and colored neighborhood $\mathcal N^{(v)}$. Then, the {simplicial susceptible-infected-susceptible (SIS)} model \citep{bruneo2012markovian}, which imposes additional higher-order terms on the infection rates of vertices, can be modeled as
\begin{align}
    \begin{split} \label{eq:infect}
        &\lambda^{S \to I}(\mathcal N^{(v)}) = \tau \graphsum{\mathcal N^{(v)}}{ \mathbbm 1_{\{I\}} } \\
        &+ \tau_\triangle \sum_{(v, n, n') \in \Delta_v}
        \left[ \mathbbm 1_{\{I\}}( \mathcal N^{(v)}[n] ) + \mathbbm 1_{\{I\}}( \mathcal N^{(v)}[n'] ) \right], 
    \end{split} \\
    &\lambda^{I \to S}(\mathcal N^{(v)}) = \gamma, \label{eq:recover}
\end{align}
for pairwise infection rate $\tau$, triangle (clique) infection rate $\tau_\triangle$, recovery rate $\gamma$ and indicator function $\mathbbm 1_A$. Here, the summation is over all unique triangles $\Delta_v$ involving $v$. This model is more realistic than the standard SIS model when shared acquaintances meet more often (see Figure~\ref{fig:explain1}, \citep{battiston2020networks}). In our experiments, we also consider the standard SIS model where $\tau_\triangle = 0$, which can also be understood as a result of microscopic contact processes \citep{G_mez_2010}, for which we similarly imagine higher-order interactions to be of interest. For a susceptible-infected-removed (SIR) model, $\lambda^{I \to S}$ is replaced by jumps to a third terminal state $R$. Finally, we consider the Ising Glauber dynamics \citep{glauber1963time} with states $\{U, D\}$ and
\begin{multline} \label{eq:ising}
    \lambda^{U \to D}(\mathcal N^{(v)}) = 1 - \lambda^{D \to U}(\mathcal N^{(v)}) \\
    = \left\{ 1 + \exp{ \left[ \frac{2J}{T} \graphsum{\mathcal N^{(v)}}{ (-1)^{\mathbbm 1_{\{D\}}} } \right] } \right\}^{-1}
\end{multline}
for interaction strength $J > 0$ and temperature $T > 0$. 

\begin{figure*}[t]
    \center
    \includegraphics[width=0.85\linewidth]{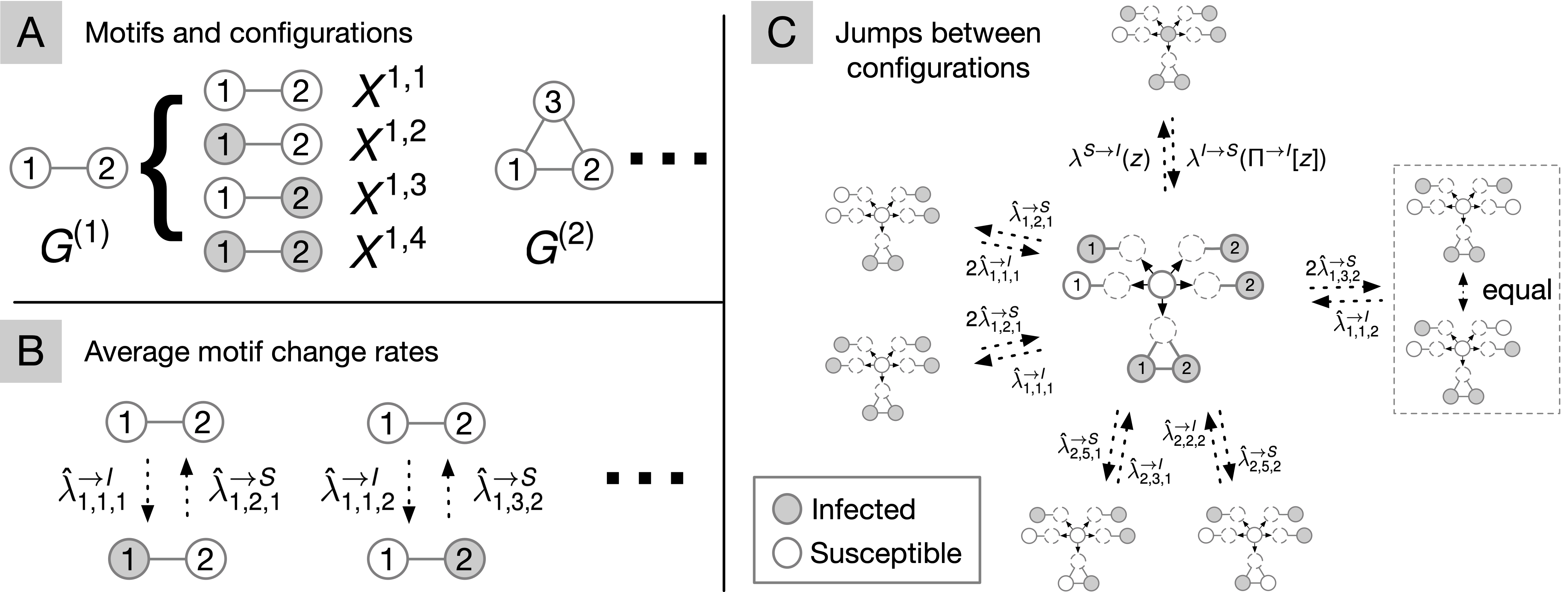}
    \caption{Schematic illustration of the MMF equations for two states. \textbf{A:} A fixed, finite number of network motifs with associated motif colorings. \textbf{B:} Common, shared jump rates between motif colorings. \textbf{C:} A visualization of all jumps from and to configuration $z$ (center), where the jump rates are given by the number of motif configurations times their shared motif jump rates. Here, $d^1 = (2,2)$, $d^2 = (0,0,1)$ and $z^{1,1} = (1,1,0,0)$, $z^{1,2} = (0,0,2,0)$, $z^{2,3} = (0,1,0,0,0,0,0,0)$.}
    \label{fig:explain3}
\end{figure*}

\paragraph{Mean-Field Approximation} 
While the exact colored graphs $G_t$ can be evolved through their probability laws or their associated operator semigroup $M$, an exact analysis is typically prohibitive due to the combinatorial state space. In the limit of large graphs ($N \to \infty$), our aim will thus be to approximate by a system of ODEs $M'$ the evolution of certain population fractions, obtained by aggregating the colored graphs via some aggregation function $\varphi$ -- e.g. densities of different colors $\varphi_x(G) \equiv \frac 1 N \sum_v \mathbbm 1_{\{x\}}(G[v])$ -- such that the diagram 
\begin{equation} \label{eq:commutative}
  \begin{tikzcd}
    G_0 \arrow{r}{M} \arrow{d}{\varphi} & G_t \arrow{d}{\varphi} \\
    \rho_0 \arrow{r}{M'} & \rho_t
  \end{tikzcd}
\end{equation}
commutes: The goal is to find a system of ODEs $M'$ that accurately models the evolution of population fractions $\rho$, such that aggregating population fractions through $\varphi$ and then applying $M'$ is equivalent to first exactly evolving the system and then aggregating.

Since the degrees are bounded, the jumps of $G_t$ are also bounded. Therefore, one expects the jumps of various $(1/N)$-scaled fractions to vanish in the limit because their quadratic variations (e.g., the running sum of squared jump sizes) vanish over finite time horizons. Consequently, even though the scaled proportions are not necessarily Markovian, their large-graph limits have continuous paths and can be described using ODEs by first performing the Doob--Meyer decomposition, which intuitively separates out a stochastic process that captures the mean of the scaled proportions and a zero-mean martingale (a stochastic process that acts like an error process or fluctuations around the mean process), and then invoking the FLLN for martingales \citep{Kurtz1981SIAM, JacodShiryaev} to claim that the fluctuations around the mean process vanish in the limit.

Denote the set of non-negative integer solutions to the Diophantine equation $y_1+y_2+\ldots+y_n=k$ by $\Theta(n,k)$. It is useful to think of $k\mapsto \Theta(n,k)$ as the equivalence class of a vector in $\mathbb{N}_0^n$ whose elements sum up to $k$ (where two vectors are equivalent if their elements have the same sum). For motifs $G^{(i)}$, consider their set of distinct colorings $\mathcal{G}^{(i)}$ and $C_i \equiv |\mathcal{G}^{(i)}| = |\mathcal X|^{N_i}$. For a vertex with hyperstub degree $d$, the possible counts of each neighboring motif coloring where the vertex participates as the $j$-th vertex role in a motif $G^{(i)}$ are elements of $\Theta(C_i, d^{i,j})$. 

Therefore, all colored neighborhoods $\mathcal N^{(v)}$ will belong to an equivalence class corresponding to a count vector (configuration) $z \in \mathcal{Z} \subset \bigtimes_{i=1}^M \bigtimes_{j=1}^{N_i} \mathbb{N}_0^{C_i}$ under an appropriate equivalence relation ${\sim}$, such that $z \equiv (z^{1,1}, \ldots, z^{1,N_1}, z^{2,1}, \ldots, z^{M,1}, \ldots, z^{M,N_M}) \in \mathcal Z$, $z^{i,j} \equiv (z^{i,j}_k, \ldots, z^{i,j}_{C_i})$, and $z^{i,j}_k \in \{0, 1, \ldots, d^{i,j}\}$ denotes the number of participations as role $j$ in neighboring motifs $G^{(i)}$ currently in the $k$-th motif coloring $X^{i, k} \in \mathcal X^{V(G^{(i)})}$. Denote the set of such equivalence classes that are compatible with $d$ and the center vertex state $x$ by $[x, d]$. To each $z \in [x, d]$ corresponds injectively some $z' = \Pi^{\to y}[z] \in [y, d]$ where the color of the center vertex is changed from $x$ to $y$, and analogously $z'' = \Pi^{\to y}_{i,k,v}[z] \in [x, d]$, where the color of a neighboring vertex that participates as role $v$ in motif $G^{(i)}$ with current motif coloring $k$, $X^{i, k} \in \mathcal X^{V(G^{(i)})}$, is changed from $X^{i, k}_v$ to $y$. Moreover, each $z \in [x, d]$ determines the colored neighborhood (up to isomorphism) of a center vertex with color $x$ and hyperstub degree $d$. 

Aggregating colored ECMs over equivalence classes from the quotient space $\mathcal{G}/{\sim}$, where $\mathcal{G}$ is the space of all colored ECMs, is tantamount to keeping track of proportions $\rho_t(x, d, z)$ of vertices in $G_t$ with color $x\in \mathcal{X}$, hyperstub degree $d$, and counts of neighboring motif colorings $z\in [x,d]$. Note that although $z$ already contains all information about $x, d$, for notational convenience we track proportions of $(x, d, z)$. As $N \to \infty$, these proportions can be described by deterministic ODEs, which we shall call the motif-based mean-field (MMF) equations.

This leads us to our main result: The MMF master equations for the limiting proportions $\rho_t$ constitute a system $M'$ of ODEs in \eqref{eq:commutative} with an accuracy going beyond existing mean field approximations, and are given by
\begin{multline} \label{eq:mmf}
    \dot{\rho_t}(x,d,z) =
    \sum_{y \in \mathcal X} \left( \Lambda^{\leftarrow y} - I \right) \rho_t(x,d,z) \lambda^{x \to y}({z}) \\
    + \sum_{y \in \mathcal X} \sum_{i,j,k,v \neq j} \left( \Lambda^{\leftarrow y}_{i,j,k,v} - I \right) \rho_t(x,d,z) z^{i,j}_k \hat \lambda^{\to y}_{i,k,v}
\end{multline}
where we aggregate rates $\lambda^{x \to y}({z})$ and $z^{i,j}_k \hat \lambda^{\to y}_{i,k,v}$ over equivalence classes corresponding to each center vertex configuration $z$ (since $z$ uniquely defines the colored neighborhood up to isomorphism) and each coloring $k$ of neighboring motifs $G^{(i)}$ respectively. Here, we defined unit operators $I$ and influx step operators $\Lambda^{\leftarrow y}, \Lambda^{\leftarrow y}_{i,j,k,v}$ acting on functions $f(x, d, z, y), f(x, d, z, y, k)$ such that we have influx by center vertex jumps from configurations $\Pi^{\to y}[z]$
\begin{align} \label{eq:game1}
     \left[ \Lambda^{\leftarrow y} f \right](x, d, z, y) = f(y, d, \Pi^{\to y}[z], x)
\end{align}
and similarly influx by jumps of all neighboring motifs' vertices that are not the center vertex $(v \neq j)$
\begin{multline} \label{eq:game2}
    \left[ \Lambda^{\leftarrow y}_{i,j,k,v} f \right](x, d, z, y, k) \\ = f(x, d, \Pi^{\to y}_{i,k,v}[z], X^{i, k}_v, \Omega^{\to y}_{i,v}[k])
\end{multline}
where $\Omega^{\to y}_{i,v}[k]$ denotes the motif coloring resulting from changing the color of vertex $v$ to $y$ in motif $i$ with coloring $k$. The jump rates of any neighbors in role $v$ of motif $G^{(i)}$ with coloring $k$ from the corresponding state $\tilde x = X^{i, k}_v$ to $y$ are approximated by the averaged jump rate over all such colored motif occurrences
\begin{equation}
    \hat \lambda^{\to y}_{i,k,v} \equiv \frac{\sum_{d, z} \rho_t(\tilde x, d, z) z^{i,v}_k \lambda^{\tilde x \to y}(z)}{\sum_{d, z} \rho_t(\tilde x, d, z) z^{i,v}_k},
\end{equation}
since a vertex in configuration $z$ participates $z^{i,v}_k$ times in the considered motif coloring. See Figure~\ref{fig:explain3} for a visualization. Finally, sampling i.i.d. initial states from some $P_0 \colon \mathcal X \to [0, 1]$, the initial conditions are given by 
\begin{multline}
    \rho_0(x, d, z) = P_0(x) P(d) \\ \cdot \prod_{i,j,k,v \neq j} \left[ P_0(X^{i, k}_v) \cdot \mathbbm 1_{\{x\}}(X^{i, k}_j) \right]^{z^{i,j}_k}
\end{multline} 
where $0^0 \equiv 1$. The fractions of vertices in any state $x$ are then given by $\rho_t(x) = \sum_{d, z} \rho_t(x, d, z)$.

\begin{figure}[b]
    \center
    \includegraphics[width=\linewidth]{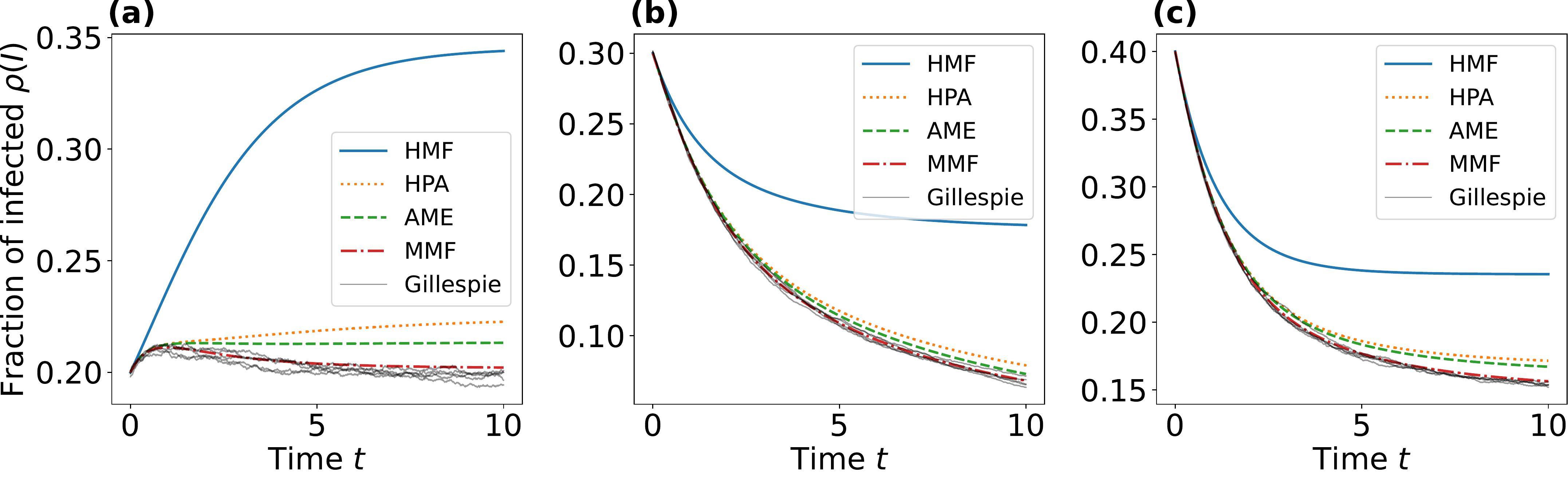}
    \includegraphics[width=\linewidth]{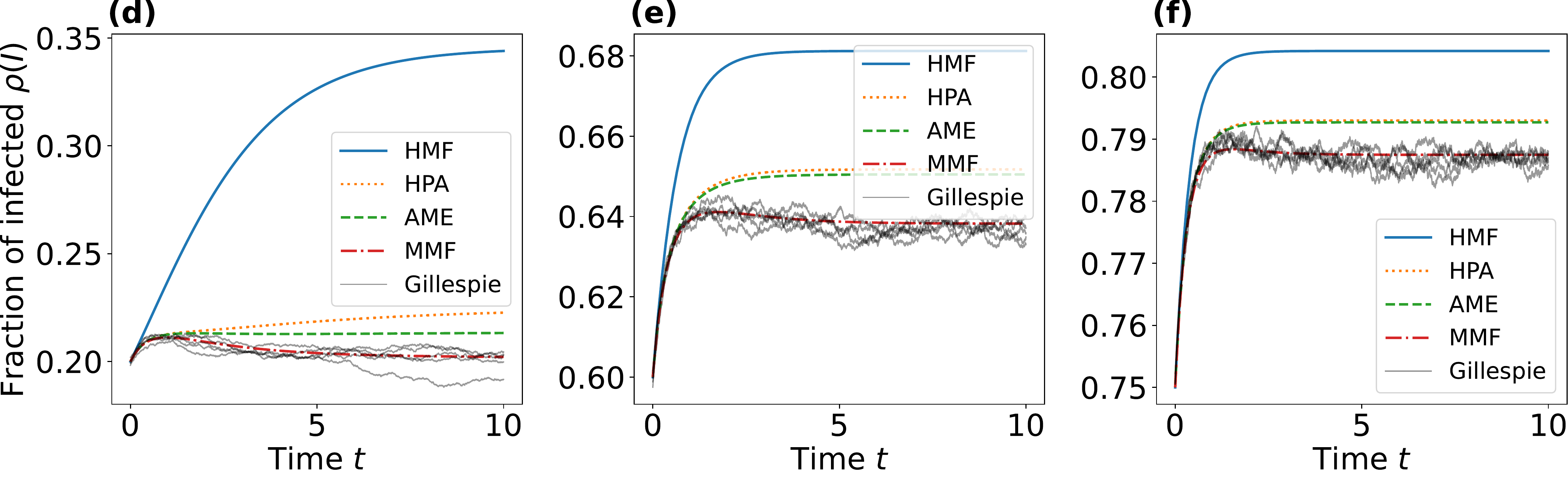}
    \caption{Mean-field approximations in the SIS model using the edge and triangle graphs as motifs. \textbf{(a-c)}: Results for $\tau = 0.3$, $\gamma = 0.9$. (a): $P^{\mathrm{a},3}$, $P_0(I) = 0.2$, (b): $P^{\mathrm{u},3}$, $P_0(I) = 0.3$, (c): $P^{\mathrm{d},2}$, $P_0(I) = 0.4$. \textbf{(d-f)}: Results for $P^{\mathrm{a},3}$. (d): $\tau = 0.3$, $\gamma = 0.9$, $P_0(I) = 0.2$, (e): $\tau = 0.5$, $\gamma = 0.7$, $P_0(I) = 0.6$, (f): $\tau = 0.65$, $\gamma = 0.55$, $P_0(I) = 0.75$.}
    \label{fig:SIS}
\end{figure}

\begin{figure}[t]
    \center
    \includegraphics[width=\linewidth]{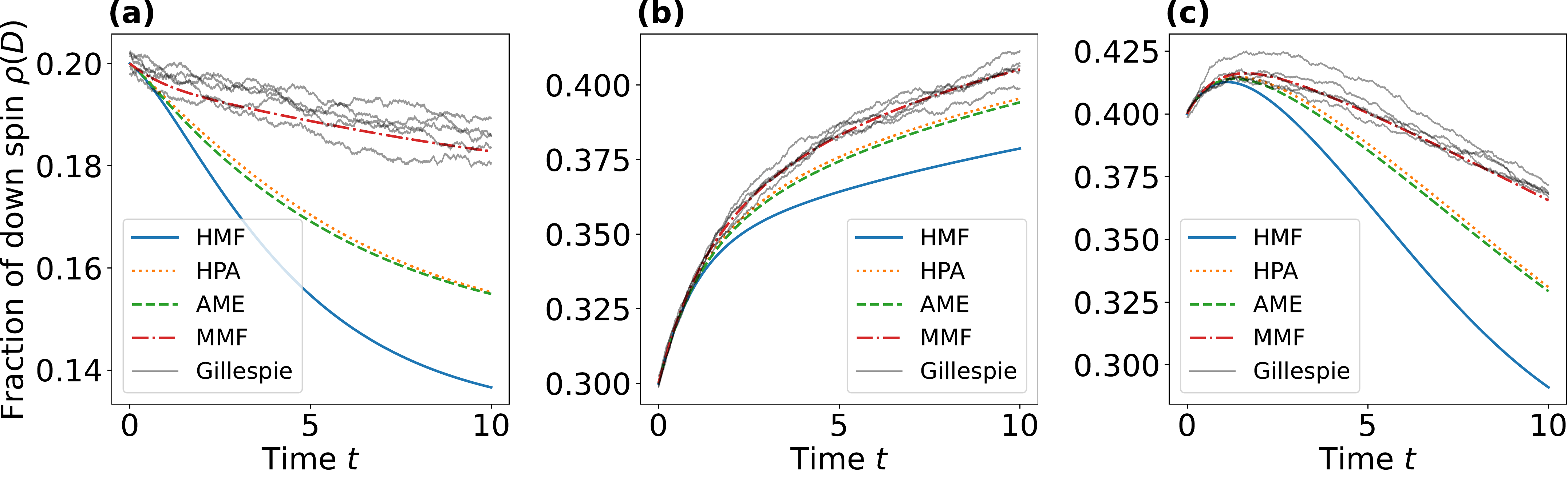}
    \includegraphics[width=\linewidth]{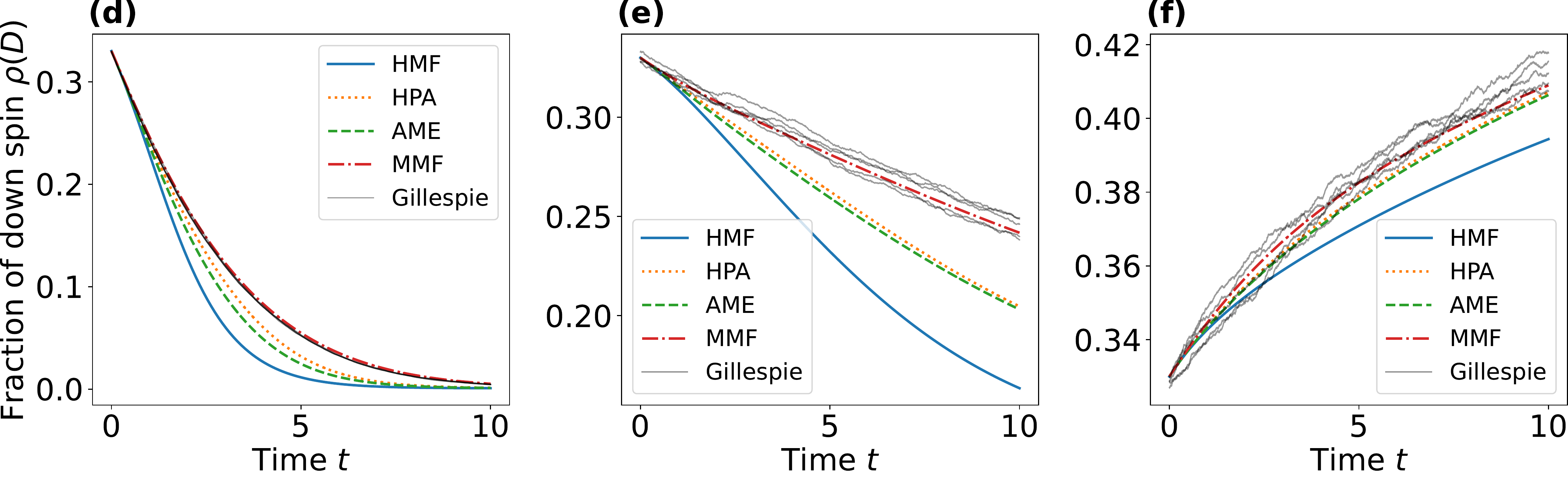}
    \caption{Mean-field approximations in the Ising Glauber model using the edge and triangle graphs as motifs. \textbf{(a-c)}: Results for $TJ^{-1}=3$. (a): $P^{\mathrm{a},3}$, $P_0(D) = 0.2$, (b):  $P^{\mathrm{u},3}$, $P_0(D) = 0.3$, (c): $P^{\mathrm{d},2}$, $P_0(D) = 0.4$. \textbf{(d-f)}: Results for $P^{\mathrm{a},3}$, $P_0(D) = 0.33$. (d): $TJ^{-1}=1$, (e): $TJ^{-1}=3$ (f): $TJ^{-1}=4$.}
    \label{fig:Ising}
\end{figure}

\begin{figure}[b]
    \center
    \includegraphics[width=\linewidth]{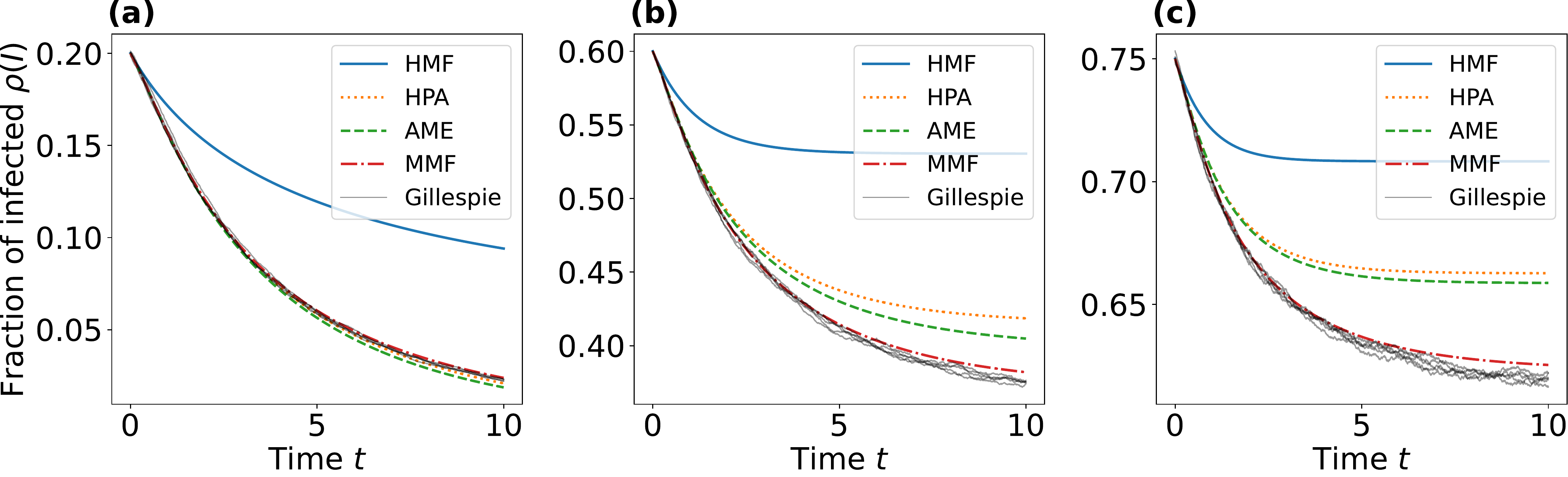}
    \caption{Mean-field approximations in the SIS model using the edge and square graphs as motifs and $P^{\mathrm{a},2}$. (a): $\tau = 0.3$, $\gamma = 0.9$, $P_0(I) = 0.2$, (b): $\tau = 0.5$, $\gamma = 0.7$, $P_0(I) = 0.6$, (c): $\tau = 0.65$, $\gamma = 0.55$, $P_0(I) = 0.75$.}
    \label{fig:Square}
\end{figure}

The biggest appeal of the MMF equations \eqref{eq:mmf} is their simplicity and intuitiveness. While they may generally not be asymptotically exact, experimentally we find that they are quite accurate. Note that as a special case, we obtain classical approximations such as AME \citep{gleeson2011high} and thereby coarser approximations \citep{gleeson2013binary} for degree distributions $\tilde P \colon \mathbb N_0 \to [0, 1]$ by considering only the edge motif $G^{(1)}$, assuming binomial role distributions and aggregating equivalent terms, i.e. $P(d^{1,1}, d^{1,2}) = \tilde P(d^{1,1} + d^{1,2}) \cdot \binom{d^{1,1} + d^{1,2}}{d^{1,1}} (\frac 1 2)^{d^{1,1}} (\frac 1 2)^{d^{1,2}}$. 

\paragraph{Numerical Evaluation} 
For numerical purposes, we generate equations only for $P$-supported hyperstub degrees $d$ and simulate rescaled proportions $\rho_t(x, z \mid d) \equiv \rho_t(x, d, z) / P(d)$. For fast ECM graph generation, we drop leftover hyperstubs (in our experiments, this amounts to less than $0.5 \%$ of all generated stubs, leading to only slight inaccuracies) instead of resampling until cardinality constraints are satisfied and allow but ignore self-loops and multi-edges. We use a third-order numerical integrator and compare MMF against the approximate master equations (AME) \citep{gleeson2011high}, the heterogeneous pair approximation (HPA) \citep{pugliese2009heterogeneous}, the heterogeneous mean-field approximation (HMF) \citep{sood2005voter} and exact Gillespie simulations on graphs of size $N=100000$. For use by the wider community, Python code is available at \citep{cui_kai_2022_5653343}.

For two given, arbitrary network motifs $G^{(1)}$, $G^{(2)}$ we consider the three parametrized families of antidiagonal, uniform and diagonal hyperstub degree distributions $P^{\mathrm{a},\theta}$, $P^{\mathrm{u},\theta}$ and $P^{\mathrm{d},\theta}$ with parameter $\theta \in \mathbb N$: For $P^{\mathrm{a},\theta}$, we put uniform mass $ 1 /{(\theta+1)}$ on each case where $\sum_j d^{1,j} = k$ and $\sum_j d^{2,j} = \theta-k$ for $k=0,1,\ldots,\theta$. In each case, we shall assume a uniform distribution over motif roles, resulting in a product of multinomials $P^{\mathrm{a},\theta}(d) \equiv  \frac{1}{ (\theta+1)} \sum_{k=0}^\theta \mathbbm 1_{\{k\}}(\sum_j d^{1,j} + d^{2,j}) \cdot \prod_{i \in \{1,2\}} \mathrm{Mult}(d^i \mid k, \frac 1 {N_i} {1_{N_i}})$, where $1_{N_i}$ is the $N_i$-dimensional one-vector. For $P^{\mathrm{u},\theta}$ and $P^{\mathrm{d},\theta}$ we similarly put equal probability mass whenever $\sum_j d^{1,j} + \sum_j d^{2,j} \leq \theta$ and $\sum_j d^{1,j} = \sum_j d^{2,j} = \theta$ respectively.

\begin{figure}
    \center
    \includegraphics[width=\linewidth]{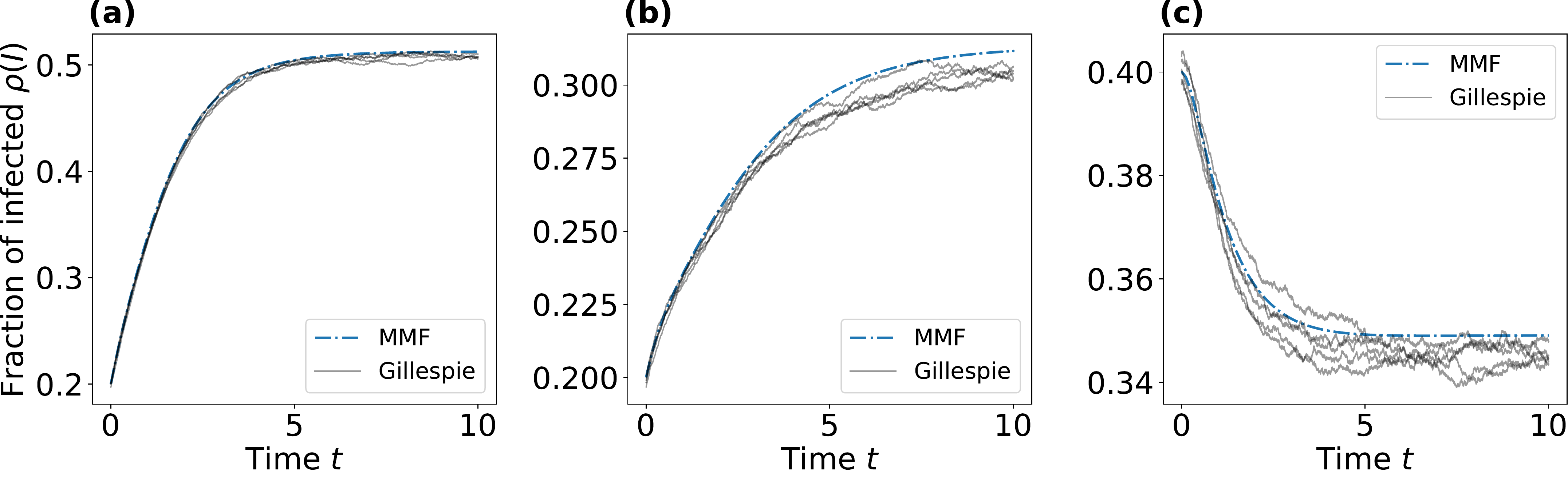}
    \caption{A comparison between MMF and numerical Gillespie simulation in the simplicial SIS model using $\tau_\triangle = \tau$, the edge and triangle graphs as motifs and $\tau = 0.3$, $\gamma = 0.9$. (a): $P^{\mathrm{a},3}$, $P_0(I) = 0.2$, (b):  $P^{\mathrm{u},3}$, $P_0(I) = 0.2$, (c): $P^{\mathrm{d},2}$, $P_0(I) = 0.4$. }
    \label{fig:Simplicial}
\end{figure}

\begin{figure}
    \center
    \includegraphics[width=\linewidth]{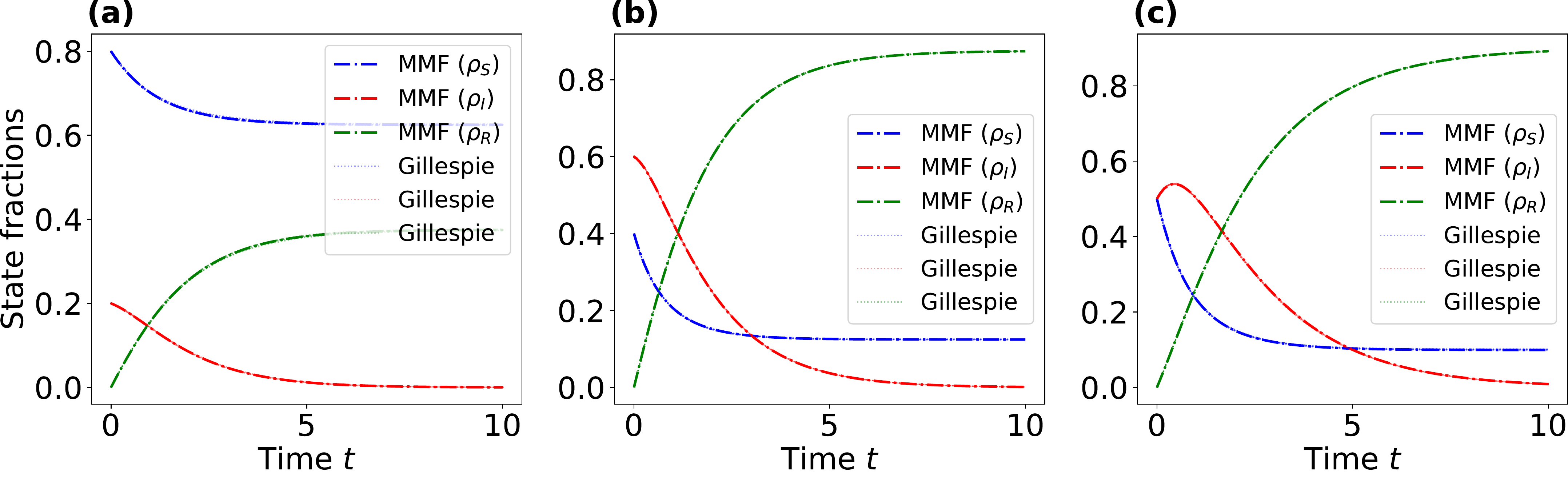}
    \caption{The MMF approximation is almost indiscernible from the numerical simulation in the SIR model using the edge and triangle graphs as motifs and $P^{\mathrm{a},2}$. (a): $\tau = 0.3$, $\gamma = 0.9$, $1-P_0(S)=P_0(I) = 0.2$, (b): $\tau = 0.5$, $\gamma = 0.7$, $1-P_0(S)=P_0(I) = 0.6$, (c): $\tau = 0.6$, $\gamma = 0.5$, $1-P_0(S)=P_0(I) = 0.5$.}
    \label{fig:SIR}
\end{figure}

On the ECM graphs with edge and triangle motifs ($G^{(1)}, G^{(2)}$ from Figure~\ref{fig:explain1}), we find that our approximation matches well with the numerical Gillespie simulation. For the SIS dynamics (\ref{eq:infect},~\ref{eq:recover}) in Figure~\ref{fig:SIS}, our approximation outperforms other approximation methods over a range of (hyperstub) degree distributions and dynamics parameters. Similar assertions hold for the Ising Glauber dynamics \eqref{eq:ising} in Figure~\ref{fig:Ising}, where existing mean-field approximations become highly inaccurate near the critical point due to the high clustering of the considered graphs. Furthermore, our approximations remain quite accurate also e.g. for graphs with edge and square motifs ($G^{(1)}, G^{(3)}$ in Figure~\ref{fig:explain1}) as seen in Figure~\ref{fig:Square}. For the simplicial version of the SIS dynamics, in Figure~\ref{fig:Simplicial} we find that the accuracy of our approximations is acceptable, while existing degree-based approximations are unable to handle simplicial dynamics by design. Finally, we verify the accuracy of our proposed framework on the SIR dynamics model in Figure~\ref{fig:SIR} with non-binary states, where the Gillespie simulation for $N=100000$ is almost indiscernible from the predicted mean-field proportions, showing the generality of our approach.

\paragraph{Discussion}
We have proposed motif-based mean-field equations for arbitrary neighborhood-dependent jump dynamics on a highly adjustable random graph model, considering both higher-order graph structures and dynamics. Numerical examples show that our approximations are quite accurate. Potential extensions include the consideration of general $k$-hop neighborhoods with $k>1$, control and lumping of equations \citep{grossmann2018lumping, khudabukhsh2019approximate} under additional assumptions on motif roles to improve tractability. Finally, for applications, estimating hyperstub degree distributions constitutes another important problem, as an identifiability problem arises from counting larger motifs that include smaller motifs.

\begin{acknowledgments}
This work has been funded by the LOEWE initiative (Hesse, Germany) within the emergenCITY center. HK acknowledges support by the German Research Foundation (DFG) via the Collaborative Research Center (CRC) 1053 – MAKI. WRK received no specific grant for this research from any funding agency in the public, commercial, or not-for-profit sectors.  
\end{acknowledgments}

\appendix
% \nocite{*}
\bibliography{root}

\end{document}